\documentclass[aps,pra,twocolumn,groupedaddress,10pt]{revtex4-1}

\usepackage{amsmath}
\usepackage{graphicx}
\usepackage{multirow}
\usepackage{bm}
\usepackage{caption}
\usepackage{array}
\usepackage{color}
\newcommand\opex{Opt. Express}

\begin{document}
\title{Bound states in the continuum on  
  periodic structures surrounded by strong resonances}

\author{Lijun Yuan}
\email{Corresponding author: ljyuan@ctbu.edu.cn}
\affiliation{College of Mathematics and Statistics, Chongqing Technology and Business University, Chongqing, 
China}
\author{Ya Yan Lu}
\affiliation{Department of Mathematics, City University of Hong Kong, Hong Kong}

\begin{abstract}
Bound states in the continuum (BICs) are trapped or guided modes with
their frequencies in the frequency intervals of the radiation modes. On periodic structures,
a BIC is surrounded by a family of resonant modes 
with their quality factors approaching infinity. Typically the quality
factors are proportional to $1/|{\bm \beta} - {\bm \beta}_*|^2$, where
${\bm \beta}$ and ${\bm \beta}_*$ are the Bloch wavevectors of the
resonant modes and the BIC, respectively. But for some special
BICs, the quality factors are proportional to $1/|{\bm \beta} - {\bm
  \beta}_*|^4$. In this paper, a general condition is derived for such
special BICs on two-dimensional periodic structures. 
As a numerical example, we use the general condition to
calculate special BICs, which are antisymmetric standing waves, on a periodic array of
circular cylinders, and show their dependence on parameters. 
 The special BICs are important for practical applications, because they produce resonances with large quality factors 
for a very large range of ${\bm \beta}$. 
\end{abstract}

\maketitle

\section{Introduction}

Bound states in the continuum (BICs), first studied by Von
Neumann and Wigner for quantum systems \cite{neumann29}, are trapped
or guided modes with their frequencies in the frequency intervals of 
radiation modes that carry power to or from infinity \cite{hsu16}. For
light waves, BICs have been analyzed and observed for many
different structures,  including waveguides with local 
distortions \cite{evans91,gold92,evans94,port98,bulg08}, waveguides with lateral leaky structures \cite{plot11,moli12,weim13,zou15}, and periodic
structures sandwiched between or surrounded by homogeneous media
\cite{bonnet94,padd00,tikh02,shipman03,shipman07,lee12,hu15,port05,mari08,ndan10,hsu13_1,hsu13_2,yang14,zhen14,bulg14b,bulg15,bulg16,li16,gan16,gao16,ni16,yuan17,bulg17,hu17josab,bulg17prl,yuan17_4,bulg17pra,hu18}. The
BICs on periodic structures are particularly interesting, because they
are surrounded by families of resonant modes (depending on the
wavevector) with quality factors ($Q$-factors) tending to infinity, and they give rise to collapsing Fano resonances
corresponding to discontinuities in the
transmission and reflection spectra \cite{shipman05,shipman12}. 
The high-$Q$ resonances and the related strong local fields
\cite{mocella15,yoon15} can be
used to enhance light-matter interactions for applications in lasing \cite{kodi17},
nonlinear optics \cite{yuan17_2}, etc. Collapsing Fano resonances can
be exploited in filtering, sensing, and switching applications
\cite{foley14,cui16}. 

If the structures are symmetric, the BICs and the radiation modes may have
incompatible symmetry so that they are automatically decoupled.  These
so-called symmetry-protected BICs are well known 
\cite{bonnet94,padd00,tikh02,shipman03,shipman07,lee12,hu15}. Their
existence 
can be rigorously proved \cite{evans94,bonnet94,shipman07,hu15}, and they
are robust against small structural perturbations that preserve the
required symmetry.   On periodic structures, there are also BICs that do not
have a symmetry mismatch with the radiation modes
\cite{port05,mari08,ndan10,hsu13_1,hsu13_2,yang14,zhen14,bulg14b,bulg15,bulg16,gan16,li16,gao16,ni16,yuan17,bulg17,hu17josab,bulg17prl,yuan17_4,bulg17pra,hu18}. These 
BICs are often considered as unprotected by symmetry, but for
some important cases, they appear to depend crucially on the
symmetry, and they continue to exist  when
geometric and material parameters are varied with the relevant
symmetries kept intact
\cite{zhen14,bulg17prl,bulg17pra,yuan17_4,hu18}. In fact, it has been 
shown that under the right
conditions, these BICs are robust
against any structural changes that preserve the relevant symmetries
\cite{yuan17_4}. 

On periodic structures,  a  BIC is a guided mode, but it belongs to a
family of resonant modes, and can be regarded as a special 
resonant mode with an infinite $Q$-factor. Let ${\bm \beta}_*$ be the Bloch
wavevector of a BIC, then there is a related family of resonant modes depending on
wavevector ${\bm \beta}$. The $Q$-factors of the resonant modes
typically satisfy $Q \sim 1/|{\bm \beta}-{\bm \beta}_*|^{2}$. Clearly, a
resonant mode with an arbitrarily large $Q$-factor can be obtained if
${\bm \beta}$ is chosen to be sufficiently close to ${\bm \beta}_*$, and
an arbitrarily large local field can be induced by an incident wave
with the wavevector ${\bm \beta}$.  However, practical
applications of the strong field enhancement can be limited by the
difficulty of controlling ${\bm \beta}$ to high precision, in addition
to other practical issues such as fabrication errors \cite{lavri17}, material
dissipation \cite{yoon15}, variations in
different periods, finite sizes \cite{bulg17_2,sadr17,tagh17}, etc. In \cite{yuan17_2}, we showed 
that for symmetric standing waves, which are BICs with ${\bm   \beta}_*
= {\bm 0}$ and are unprotected by symmetry in 
the usual sense, the $Q$-factors of the associated resonant modes satisfy an inverse
fourth power asymptotic relation, i.e., $Q \sim 1/|{\bm
  \beta} - {\bm \beta}_*|^{4}$. In that case, resonances with large
$Q$-factors and strong local fields can be realized with a
much more relaxed condition on ${\bm \beta}$. This property has been
used to show that optical bistability can be induced by a very
weak incident wave \cite{yuan17_2}, and  it should be useful in other
applications that require a significant field enhancement.
In general, resonant modes near antisymmetric
standing waves (ASWs), which are symmetry-protected BICs with ${\bm
  \beta}_*={\bm 0}$, only satisfy the
inverse quadratic asymptotic relation, but Bulgakov
and Maksimov found a few examples for which the
inverse fourth power relation is satisfied \cite{bulg17_2}. 

In this paper, we derive a general condition for those special BICs with the
$Q$-factors of the associated resonant modes satisfying the inverse
fourth power relation. For 
simplicity, the theory is developed for two-dimensional (2D)
periodic structures. We use a perturbation method assuming $|{\bm
  \beta}-{\bm \beta}_*|$ is small. The condition is given in  
integrals involving the BIC itself and related diffraction solutions 
for incident waves with the same frequency
and same wavevector.  With this general condition, it becomes feasible  to
systematically search the parameter values of the periodic structure
supporting the special BICs. As numerical examples, we
calculate special BICs on a periodic array of circular dielectric
cylinders, and show their dependence on the parameters. 

\section{BICs and resonant modes}

We consider 2D dielectric structures which are invariant in $z$, periodic in $y$ with
period $L$, and bounded in the $x$ direction by $|x|<D$ for some
constant $D$, where $\{x,y,z\}$ is a Cartesian coordinate system. The
surrounding medium for $|x| > D$ is assumed to be vacuum. Therefore, the dielectric
function $\epsilon$ satisfies $\epsilon(x,y+L)=\epsilon(x,y)$ for all
$(x,y)$, and
$\epsilon(x,y)=1$ for $|x| > D$. For the $E$ polarization, 
the $z$-component of the electric field, denoted as $u$, satisfies the
Helmholtz  equation 
\begin{equation}
\label{eq:helm} \frac{\partial^2 u}{\partial x^2} + \frac{\partial^2
  u}{\partial y^2} + k^2 \epsilon u = 0, 
\end{equation}
where $k = \omega/c$ is the free space wavenumber, $\omega$ is the
angular frequency, $c$ is the speed of light in vacuum, and the
time dependence is assumed to be $e^{- i \omega t}$. 

A Bloch mode on the periodic structure is a solution of Eq.~(\ref{eq:helm}) given as
\begin{equation}
\label{eq:bloch} u = \phi(x,y) e^{i \beta y}, 
\end{equation}
where $\phi$ is periodic in $y$ with period $L$ and $\beta$ is the
real 
Bloch wavenumber. Due to the periodicity of $\phi$, $\beta$ can be
restricted to the interval $[-\pi/L, \pi/  L]$. 
If $\phi \to 0$ as $|x| \to \infty$, then $u$ given in
Eq.~(\ref{eq:bloch}) is a guided mode. 
Typically, guided modes that depend 
on $\beta$ and $\omega$ continuously  can only be found below the
light line, i.e., for $ k < |\beta|$. A BIC is a special guided mode above the light line, i.e., $\beta$
and $k$ satisfy the condition $k> |\beta|$. 
For a given structure, BICs can only exist at isolated points in the
$\beta\omega$ plane. 

In the homogeneous media given by $|x| > D$, we can expand a Bloch mode in
plane waves, that is 
\begin{equation}
\label{planewave} u(x,y) = \sum_{j = - \infty}
^{\infty}c_{ j}^{\pm} e^{  i (\beta_j y \pm  \alpha_ j x)}
\end{equation}
where the ``$+$'' and ``$-$'' signs are chosen 
for $x > D$ and $x < -D$ respectively, and 
\begin{equation}
  \label{defab}
\beta_j = \beta + \frac{2\pi j}{L}, \quad 
\alpha_j = \sqrt{k^2 - \beta_j^2}.  
\end{equation}
If $k < |\beta_j|$, then $\alpha_j = i \sqrt{\beta_j^2 - k^2}$ is pure
imaginary, and the corresponding plane wave is evanescent.
 For a BIC, one or more $\alpha_j$ are real, then the
corresponding coefficients $c_j^\pm$ must vanish, since the BIC must
decay to zero as $|x| \to \infty$. 

Above the light line, if the frequency  $\omega$ is allowed to be
complex, there are Bloch mode solutions that depend
on a real wavenumber $\beta$ continuously.  These solutions are the resonant
modes, and they satisfy outgoing radiation conditions
as $x \to \pm \infty$. Due to the time 
dependence $e^{-i \omega t}$, the imaginary part of the complex frequency of a resonant
mode must be negative, so that its amplitude decays with time.  The
$Q$-factor is given by $Q = - 0.5
\mbox{Re}(\omega)/ \mbox{Im}(\omega)$. The expansion (\ref{planewave})
is still valid, but the complex square root for $\alpha_j$ must be defined
to maintain continuity as $\mbox{Im}(\omega) \to 0$. This can be 
achieved by using a square root with a branch cut along the negative
imaginary axis (instead of the negative real axis), that is, if $\xi = |\xi| e^{ i \theta}$ for $-\pi/2 <
\theta \le 3\pi/2$, then $\sqrt{\xi} = \sqrt{|\xi|} e^{i
  \theta/2}$.
 Notice that $\alpha_0$ and probably a few other
$\alpha_j$ have negative real parts. Therefore, a resonant mode blows
up as $|x| \to \infty$.  As $\beta$ is continuously varied following a family of
resonant modes, $\mbox{Im}(\omega)$ may become zero at some special
values of $\beta$, and they correspond to the BICs. Therefore, although a BIC is a
guided mode, it belongs to a family of resonant modes, and it can
be regarded as a special resonant mode with an infinite $Q$-factor. 

\section{Perturbation  analysis} 

Given a BIC on a periodic structure with frequency $\omega_*$ and Bloch
wavenumber $\beta_*$, we are interested in the complex frequency
$\omega$ and the $Q$-factor of the nearby resonant mode for wavenumber
$\beta$ close to $\beta_*$. 
For simplicity, we 
scale the BIC such that 
\begin{equation}
  \frac{1}{L^2} \int_\Omega |\phi_*|^2 d{\bm r} = 1, 
\end{equation}
where $\Omega$ is one period of the structure given by $ -L/2 < y <
L/2$ and $-\infty < x < \infty$. We also assume $k_* =\omega_*/c$
satisfies the following condition 
\begin{equation}
  \label{cutoff1}
|\beta_*| < k_* < \frac{2\pi}{L}- |\beta_*|.
\end{equation}
This implies that $\alpha_{*0}$ (also denoted as $\alpha_*$ below) is positive and all
$\alpha_{*j}$ for $j\ne 0$ are pure imaginary, where $\alpha_{*0}$ and
$\alpha_{*j}$ are defined as in Eq.~(\ref{defab}) with $k$ and $\beta$ replaced by
$k_*$ and $\beta_*$, respectively. To analyze this problem, we use a
perturbation method assuming $\delta = \beta - \beta_*$ is small. 

Let $u_* = \phi_* e^{i \beta_* y}$ and $u=\phi e^{i \beta y}$ be the
BIC and the nearby resonant mode, respectively,  we expand $\omega$ and
$\phi$ as
\begin{eqnarray}
\label{omg_exp}
\omega &=& \omega_* + \omega_1 \delta + \omega_2 \delta^2 + \omega_3
           \delta^3 + \omega_4 \delta^4 + \ldots \\
\label{phi_exp} 
\phi &=& \phi_{\ast}  + \phi_1 \delta + \phi_2 \delta^2 + 
         \phi_3 \delta^3 + \phi_4 \delta^4 + \ldots
\end{eqnarray}
In terms of the periodic function $\phi$ given in Eq.~(\ref{eq:bloch}), the 
Helmholtz equation becomes 
\begin{equation}  
\label{eq:phi}
\frac{\partial^2 \phi}{\partial x^2} + \frac{\partial^2 \phi}{\partial 
  y^2}  + 2 i \beta \frac{\partial \phi}{\partial y} + \left( k^2 
  \epsilon - \beta^2 \right) \phi = 0. 
\end{equation} 
Inserting Eqs.~(\ref{omg_exp})-(\ref{phi_exp}) into 
Eq.~(\ref{eq:phi}),  and comparing terms of equal powers of 
$\delta$, we obtain 
\begin{eqnarray}
  \label{eq:phistar}
&&  {\cal L} \phi_* = 0, \\
\label{eq:phione}
&& {\cal L} \phi_1 =
 - 2i \partial_y \phi_* + 2(\beta_* - k_* k_1 \epsilon ) \phi_*, \\
\nonumber 
&&  {\cal L} \phi_2 = - 2i \partial_y \phi_1 + 2 (\beta_* - k_* k_1 \epsilon) 
\phi_1 \\ 
\label{eq:phitwo}
&& \hspace{1cm} + (1- k_1^2 \epsilon -2 k_* k_2 \epsilon) \phi_*, 
\end{eqnarray}
where $k_j = \omega_j/c$ for $j\ge 1$, and 
\begin{equation}
  \label{defL}
{\cal L} = \partial_x^2 + \partial_y^2 + 2i \beta_* \partial_y 
+ k_*^2  \epsilon-\beta_*^2.  
\end{equation}
In addition, $\phi_j$ must satisfy proper outgoing radiation 
conditions as $|x| \to \infty$.  

Equation (\ref{eq:phistar}) is simply the governing Helmholtz
equation of the BIC. The first order term $\phi_1$ satisfies the
inhomogeneous Eq.~(\ref{eq:phione}) which is singular and has no
solution, unless the right hand side is orthogonal to
$\phi_*$. Multiplying $\overline{\phi}_*$ (the complex conjugate of
$\phi_*$) to both sides of Eq.~(\ref{eq:phione}) and integrating on
domain $\Omega$, we obtain 
\begin{equation}
  \label{eq:k1}
k_1 = \frac{\omega_1}{c} = \frac{ \beta_* \int_\Omega |\phi_*|^2 d{\bm r} -  i 
  \int_\Omega \overline{\phi}_* \partial_y \phi_* \, d{\bm r} }
{k_{\ast} \int_\Omega \epsilon |\phi_*|^2 \, d{\bm r} }.
\end{equation}
It is easy to show that $k_1$ is real. Therefore, in general 
 $\mbox{Im}(\omega)$ is proportional to
$(\beta - \beta_*)^2$. 

For $k_1$ given above, Eq.~(\ref{eq:phione}) has a solution. 
 Similar to the plane wave expansion (\ref{planewave}), 
$\phi_1$ can be written down explicitly for $|x| > D$. 
Importantly, $\phi_1$ contains only a single
outgoing plane wave as $x \to \pm \infty$, that is 
\begin{equation}
\label{phi1_exp}
\phi_1 \sim  d_0^{\pm} e^{\pm i \alpha_{\ast  } x}, \quad x \to \pm
\infty, 
\end{equation}
where $d_0^{\pm}$ are unknown coefficients and $\alpha_* = \sqrt{k_*^2
  - \beta_*^2}$. 
A formula for $k_2$ can be derived from the solvability condition of
Eq.~(\ref{eq:phitwo}). In particular, the imaginary part of $k_2$ has
the following simple formula
\begin{equation}
  \label{eq:k2_imag}
  \mbox{Im}(k_2) 
  = \frac{\mbox{Im}(\omega_2)}{c} 
= - \frac{L \alpha_{\ast}  \left( |d_{0}^+|^2 +|d_{0}^-|^2 \right)} {2 
  k_{\ast}  \int_\Omega \epsilon |\phi_*|^2 \, d{\bm r}}.
\end{equation}
A special case of Eq.~(\ref{eq:k2_imag}) was previously derived in 
\cite{yuan17_2}. Additional details on the derivation of
Eqs.~(\ref{eq:k1}) and (\ref{eq:k2_imag}) are given in 
Appendix. 

Notice that if $\phi_1$ radiates power to $x=\pm \infty$, $d_0^+$ and
$d_0^-$ are nonzero, then $\mbox{Im}(\omega_2) \ne 0$. In that case, the
imaginary part of the complex frequency satisfies 
\begin{equation}
  \label{imgomg2}
  \mbox{Im}(\omega) \sim 
 - \frac{ c L \alpha_{\ast}  \left( |d_{0}^+|^2 +|d_{0}^-|^2 \right)} {2 
  k_{\ast}  \int_\Omega \epsilon |\phi_*|^2 \, d{\bm r}} (\beta-\beta_*)^2, 
\end{equation}
and the $Q$-factor satisfies 
\begin{equation}
  \label{Qorder2}
  Q \sim  \frac {k_{\ast}^2  \int_\Omega \epsilon |\phi_*|^2 \, d{\bm r}}
  { L \alpha_{\ast}  \left( |d_{0}^+|^2 +|d_{0}^-|^2 \right)}   (\beta-\beta_*)^{-2}.
\end{equation}
On the other hand, if $\phi_1$ does not radiate power to infinity,
then $d_0^\pm = 0$, $\mbox{Im}(\omega_2)=0$, and Eqs.~(\ref{imgomg2})
and 
(\ref{Qorder2}) are no longer valid. In that case,
$\mbox{Im}(\omega_3)$ must also be 
zero, since otherwise, $\mbox{Im}(\omega)$ changes signs when $\beta$
passes through $\beta_*$. This is not possible, since
$\mbox{Im}(\omega)$ of a resonant mode is always negative. Therefore,
if $\phi_1$ is non-radiative, we expect $\mbox{Im}(\omega) \sim
(\beta-\beta_*)^4$ and $Q \sim (\beta-\beta_*)^{-4}$.

\section{Strong resonances}
\label{sec:condition}

On periodic structures, the $Q$-factors of resonant modes around
certain special BICs satisfy an 
inverse fourth power asymptotic relation $Q \sim
(\beta-\beta_*)^{-4}$. This happens when the first order perturbation
$\phi_1$ does not radiate power to infinity, i.e., $d_0^\pm =
0$. However, to check this condition, it is necessary to solve
$\phi_1$ from Eq.~(\ref{eq:phione}). This is not very
convenient. Ideally, one would like to have a condition that involves
the BIC $\phi_*$ only. This does not seem to be possible. In the
following, we derive a condition that involves the BIC $\phi_*$ and 
related diffraction solutions for incident waves with the same
$\omega_*$ and $\beta_*$ as the BIC. 

For Eq.~(\ref{eq:helm}) with $k$ replaced
by $k_*$, we consider two diffraction problems with incident waves 
$e^{ i(\beta_* y + \alpha_* x)}$ and $e^{i (\beta_* y - \alpha_* x)}$
given in the left and right homogeneous media, respectively. The
solutions of these two diffraction problems are denoted as $u_l$ and
$u_r$, respectively, and they satisfy
\begin{equation}
  \label{defpsi}
u_j  = \varphi_j(x,y) e^{i \beta_* y}, \quad j \in \{ l, r\}, 
\end{equation}
where $\varphi_l$ and $\varphi_r$ are periodic in $y$ with
period $L$.
It should be pointed out that the existence of a BIC 
implies that the corresponding diffraction problems have no uniqueness 
\cite{bonnet94,shipman07}, but the diffraction solutions are
uniquely defined in the far field as $|x| \to \infty$.
In fact, $\varphi_l$ and $\varphi_r$ have the following
asymptotic formulae
\begin{eqnarray}
\label{leftasym} \varphi_l  &\sim& 
\begin{cases}
    e^{ i \alpha_\ast x} + R_l e^{-i 
    \alpha_\ast x},   & x\to -\infty \\
  T_l e^{i \alpha_{\ast} x },  & x \to +\infty, 
\end{cases} \cr 
\varphi_r  &\sim & 
\begin{cases}
 T_r e^{ -i \alpha_\ast x},  & x\to -\infty \\
e^{-i \alpha_* x} + R_r e^{i \alpha_* x},  & x \to +\infty,
\end{cases}
\end{eqnarray}
where $R_l$, $T_l$, $R_r$ and $T_r$ are the reflection and
transmission amplitudes associated with the left and right incident
waves, respectively. It is well known that the scattering matrix 
$
{\bm S} = \left [\begin{matrix} R_l & T_r \cr T_l & R_r \end{matrix} \right]
$
is unitary. Notice that $u_l$ and $u_r$ are easier to solve than
$\phi_1$, since they satisfy a homogeneous Helmholtz equation with a
zero right hand side. 

Equation (\ref{eq:phione}) for $\phi_1$ can be written as ${\cal L}
\phi_1 = 2G$, where 
\begin{equation}
  \label{defG}
G=- i \partial_y \phi_* + (\beta_* - k_* k_1 
\epsilon) \phi_*. 
\end{equation}
 Since $G \to 0$ exponentially as $x \to \pm \infty$, the
following integrals
\begin{equation}
  \label{defFF}
F_j = \int_\Omega \overline{\varphi}_j G d{\bm r}, \quad j \in \{l, r\}
\end{equation}
are well defined. On the other hand, $\varphi_j$ and $\phi_1$ (in general) do not
decay to zero  as $|x|
\to \infty$, it is not immediately clear whether $\overline{\varphi}_j
{\cal L} \phi_1$ is integrable on the unbounded domain
$\Omega$. However, for any $h \ge D$, we can define a rectangular domain
$\Omega_h$ given by $|y| < L/2$ and $|x| < h$, and evaluate the
integral on $\Omega_h$, then take the limit as $h \to \infty$.
Clearly, the limit must exist and 
\[
\lim_{h \to \infty} \int_{\Omega_h} \overline{\varphi}_j {\cal L} \phi_1 
d{\bm r} 
= 2 F_j, \quad j \in \{l, r\}. 
\]
In Appendix, we show that 
\begin{eqnarray}
\label{leftinc}
\lim_{h \to \infty} \int_{\Omega_h} \overline{\varphi}_l {\cal L} \phi_1 
d{\bm r} = 
2 i L \alpha_* \left( d_0^+ \overline{T}_l + d_0^-   \overline{R}_l \right), \\
\label{rightinc}
 \lim_{h \to \infty} \int_{\Omega_h} \overline{\varphi}_r {\cal L} \phi_1 
d{\bm r} = 2 i L \alpha_* \left( d_0^+ \overline{R}_r + d_0^-
  \overline{T}_r \right).  
\end{eqnarray}
Therefore,
\begin{eqnarray*}
   F_l =  i L \alpha_* \left( d_0^+ \overline{T}_l + d_0^-   \overline{R}_l \right), \\  
   F_r =    i L \alpha_* \left( d_0^+ \overline{R}_r + d_0^-
  \overline{T}_r \right).  
\end{eqnarray*}
Using the unitarity of the scattering matrix ${\bm S}$, it is easy to
show that  
\begin{equation}
|d_0^+|^2 + |d_0^-|^2  = \frac{|F_l|^2 + |F_r|^2}{ (L\alpha_{\ast})^2}.  
\end{equation}
If $(F_l,F_r) \ne (0,0)$, Eq.~(\ref{imgomg2}) can be written as 
\begin{equation}
  \label{imgomg3}
  \mbox{Im}(\omega) \sim 
 - \frac{ c \left( |F_l|^2 +|F_r|^2 \right)} {2 L \alpha_*
  k_{\ast}  \int_\Omega \epsilon |\phi_*|^2 \, d{\bm r}} (\beta-\beta_*)^2.
\end{equation}
Clearly, the condition $d_0^+ = d_0^- = 0$ is equivalent to 
\begin{equation}
\label{condFlFr}
F_l = F_r = 0.
\end{equation}

BICs are most easily found on structures with suitable symmetries. If
the structure has a reflection symmetry in the $y$ direction, it is
often possible to find ASWs which are
symmetry-protected BICs with $\beta_*=0$. Assuming the origin is chosen so that 
$\epsilon(x,y)=\epsilon(x,-y)$, then the ASWs
are odd functions of $y$. From Eq.~(\ref{eq:k1}), it is easy to see
that $k_1=0$, thus $G = -i \partial_y \phi_*$ and 
\begin{equation}
  \label{defFFs}
F_j = -i \int_\Omega \overline{\varphi}_j  \frac{\partial \phi_*}{\partial
  y} d{\bm r}, \quad j \in \{l, r\}.
\end{equation}
Notice that symmetric standing waves (which are even functions of $y$)
may also 
exist on periodic structures with a reflection symmetry in $y$. In
\cite{yuan17_2}, it is shown that $d_0^\pm = 0$ is always true for the
symmetric standing waves. This is so, because $k_1=0$ and 
$G = -i \partial_y \phi_*$ are still valid, thus $G$ is odd in
$y$. Meanwhile, $\varphi_l$ and $\varphi_r$ are even in 
$y$. Therefore, $F_l=F_r=0$.  

Propagating BICs (with $\beta_*\ne 0$) are often found on structures with an
additional reflection symmetry in the $x$ direction. With a properly
chosen origin, the dielectric function satisfies
\begin{equation}
  \label{reflection_xy}
\epsilon(x,y)=\epsilon(x,-y)=\epsilon(-x,y)  
\end{equation}
for all $(x,y)$. In that case, we can reduce the condition $F_l = F_r= 0$ 
to a single real condition. In \cite{yuan17_4}, it is shown that if
the BIC $u_*=\phi_* e^{i \beta_*  y}$ is a single mode, then it is
either even in $x$ or odd in $x$, and it  can be scaled to satisfy the
${\cal PT}$-symmetric condition 
\begin{equation}
  \label{defPT}
u_*(x,y)= \overline{u}_*(x,-y).  
\end{equation}
In particular, the ASWs should be scaled as pure imaginary functions.

It is also shown in \cite{yuan17_4} that there is a complex number
$C$ with unit magnitude, such that $u_e = C (u_l+u_r)$ and $u_o = C
(u_l-u_r)$ are even and odd in $x$, respectively, and are also ${\cal
  PT}$-symmetric. As in Eq.~(\ref{defpsi}), we associate two periodic
functions $\varphi_e$ and $\varphi_o$ with $u_e$ and $u_o$,
respectively. It is easy to see that  $\phi_*$, $\varphi_e$,
$\varphi_o$ and $G$ given in Eq.~(\ref{defG}) are all ${\cal
  PT}$-symmetric. 
Furthermore, let $F_e$ and $F_o$ be defined as in 
Eq.~(\ref{defFF}) for $j \in \{e, o\}$, then 
$F_e = C (F_l + F_r)$ and $F_o = C (F_l - F_r)$. This leads to 
\begin{equation}
|F_e|^2 + |F_o|^2 = 2 ( |F_l|^2 + |F_r|^2).
\end{equation}
If $(F_e,F_o) \ne (0,0)$, Eq.~(\ref{imgomg3}) can be written as 
\begin{equation}
  \label{imgomg4}
  \mbox{Im}(\omega) \sim 
 - \frac{ c \left( |F_e|^2 +|F_o|^2 \right)} {4 L \alpha_*
  k_{\ast}  \int_\Omega \epsilon |\phi_*|^2 \, d{\bm r}} (\beta-\beta_*)^2.
\end{equation}
Clearly, the condition $F_l=F_r=0$ is equivalent to 
\begin{equation}
  \label{condFeFo}
F_e = F_o = 0.
\end{equation}
If a function satisfies the ${\cal PT}$-symmetric condition
(\ref{defPT}), its real  part is even in $y$ and its imaginary part is
odd in $y$. Therefore, $F_e$ and $F_o$ are always real. 
If the BIC is even in $x$, then $F_o$ is always zero, and it is
only necessary to check one real condition 
$F_e=0$. Similarly, if the BIC is odd in $x$, the only condition is
$F_o=0$. 
For ASWs on periodic structures with the double reflection symmetry
(\ref{reflection_xy}), $G=-i \partial_y \phi_*$ is real and even in $y$, and the
corresponding diffraction solutions $u_e$ and $u_o$ are also real even 
functions of $y$.

\section{Numerical examples}
\label{sec:example}

\begin{figure}[htb]
\centering 
\includegraphics[scale=0.7]{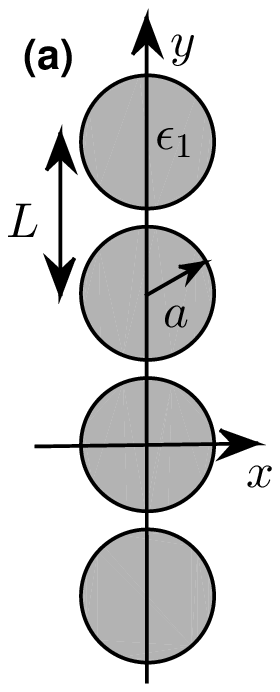} \hspace{0.3cm}
\includegraphics[scale=0.75]{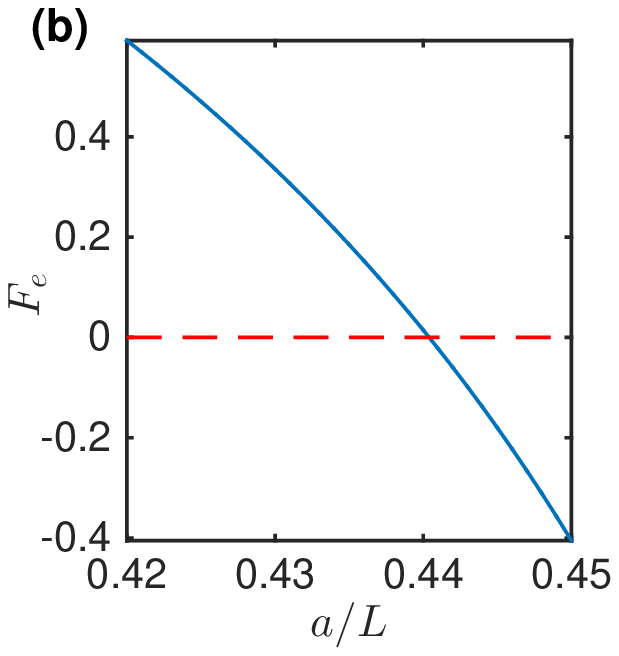}
\caption{(a): A periodic array of circular cylinders surrounded by
  air. (b) $F_e$ of the first ASW as a function of radius $a$ 
  for $\epsilon_1 = 8.2$.}
\label{figone}
\end{figure}

In this section, some numerical examples are presented to  validate
and illustrate the theoretical results developed in the previous sections. 
As shown in Fig.~\ref{figone}(a), 
we consider a single periodic array of dielectric circular cylinders  
surrounded by air. The radius and dielectric constant of the cylinders
are $a$ and $\epsilon_1$, respectively. 
BICs on such a periodic array have been extensively investigated
before \cite{shipman03,hu15,bulg14b,yuan17}. For $a = 0.4 L$,
$\epsilon_1 = 8.2$ and the $E$ polarization, 
the array supports five ASWs and one propagating BIC. The frequencies
and Bloch 
wavenumbers of these BICs are listed in the first and second columns
of Table~\ref{table1}, respectively. 
\begin{table}[htp]
\centering 
\begin{tabular}{c|c||c|c} \hline 
$\omega_* L/(2\pi c)$  &  $\beta_* L/(2\pi)$ & $a_2$: Eq.~(\ref{eq:a2})
  & $a_2$: approximation \\ \hline 
$0.4104$  & $0$ & $0.0325$  & $0.0325$\\ \hline
$0.5458$ & $0$ &   $0.1330$   & $0.1331$   \\ \hline
$0.7087$ & $0$ & $ 0.1429$ & $0.1428$ \\ \hline 
$0.8080$ & $0$ & $0.0824$   &  $0.0825$\\ \hline
$0.8749$ & $0$ & $ 0.0766$ & $0.0769$  \\ \hline
$0.6872$ & $0.2038$ & $0.1033$ & $0.1040$ \\ \hline
\end{tabular}
\caption{Frequencies and Bloch wavenumbers of six BICs on a periodic
  array of circular cylinders  with radius $a=0.4L$ and dielectric
  constant $\epsilon_1=8.2$, and their exact and approximate
  coefficients $a_2$.}
\label{table1} 
\end{table}

First, we check the formula for $\mbox{Im}(\omega)$ for ordinary BICs
where $\phi_1$ radiates power to infinity. The periodic array has
reflection symmetries in both $x$ and $y$ directions, thus,
Eq.~(\ref{imgomg4}) is applicable. In terms of the 
normalized frequency and normalized  wavenumber, Eq.~(\ref{imgomg4}) can be
written as 
\begin{equation}
  \label{imgomg5}
\frac{ \mbox{Im}(\omega) L} {2\pi c}
\sim -a_2 \left[ \frac{(\beta-\beta_*)L}{2\pi}\right]^2, 
\end{equation}
where $a_2$ is a dimensionless coefficient given by 
\begin{equation}
  \label{eq:a2}
a_2 = \frac{ \pi \left( |F_e|^2 + |F_o|^2 \right)}{ 2L^2 k_* \alpha_*
  \int \epsilon |\phi_*|^2 d{\bm r} }.  
\end{equation}
Recall that $F_e$ and $F_o$ are real, and one of them is always zero. 
For each BIC listed in Table~\ref{table1}, we calculate $a_2$ by
Eq.~(\ref{eq:a2}), and also find an approximation of $a_2$ by a
quadratic polynomial fitting the numerical values of 
$\mbox{Im}(\omega)$ for $\beta = \beta_*$ and $\beta_* \pm
0.02\pi/L$.  As shown in the third and fourth columns
of Table~\ref{table1}, the exact and approximate values of $a_2$ agree
very well. This confirms that Eqs.~(\ref{imgomg5}) and (\ref{eq:a2})
are correct.

\begin{figure}[htb]
\centering 
\includegraphics[scale=0.5]{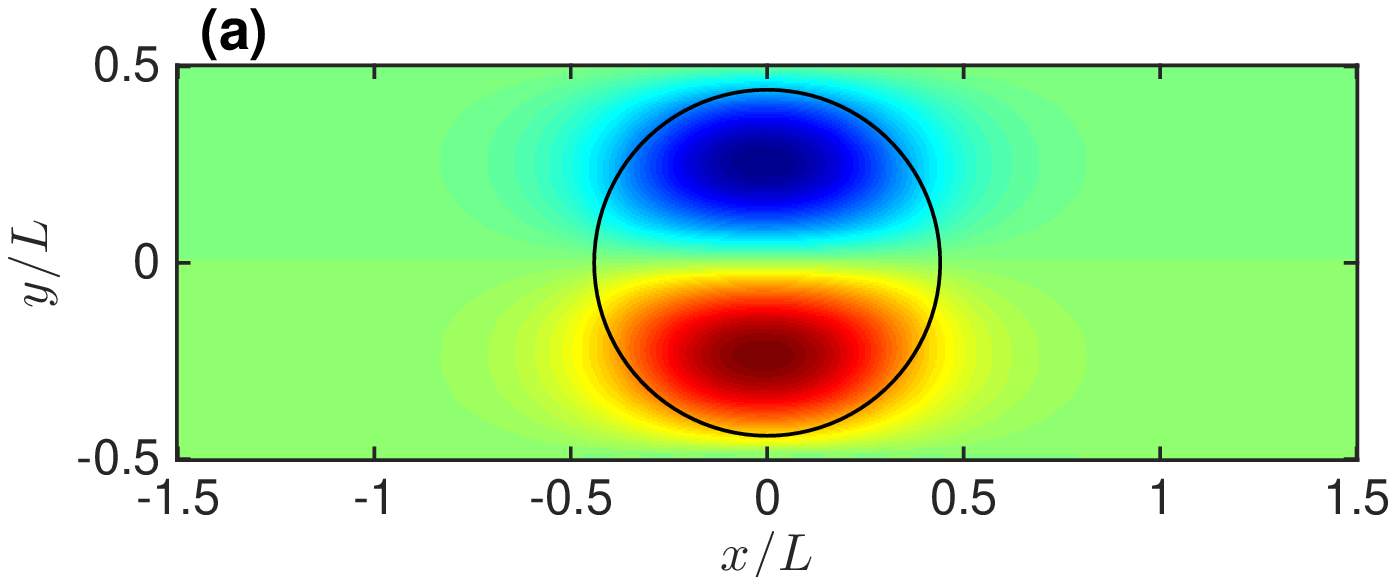}
\includegraphics[scale=0.5]{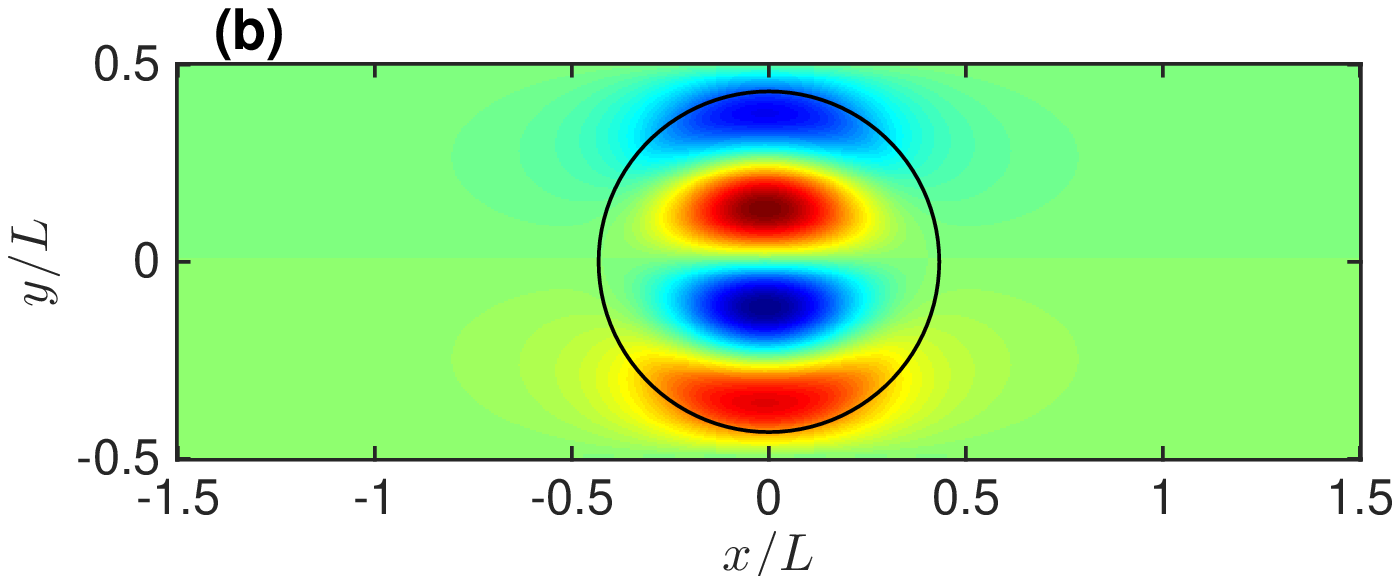}
\caption{Wave field patterns of two special $x$-even ASWs on
  periodic arrays of circular 
  cylinders with $\epsilon_1=8.2$. (a) The first ASW for
  $a=0.4404L$. (b) The fifth ASW for $a=0.4323L$.} 
\label{fig_two}
\end{figure}

We are interested in the special BICs surrounded by
strong resonances with $Q \sim 1/(\beta-\beta_*)^4$. It is
known that the 
symmetric standing waves (even in $y$) are examples of such special
BICs \cite{yuan17_2}, and they exist when $a$ and $\epsilon_1$ lie on
a curve in the $a\epsilon_1$ plane \cite{yuan17}. Bulgakov and
Maksimov \cite{bulg17_2} found a number of ASWs which also have this
special property.  
Using the perturbation theory developed in previous sections, we
can find these special BICs systematically by searching the parameters $a$ and
$\epsilon_1$, such that $F_e=0$ for an $x$-even BIC or $F_o=0$ for an
$x$-odd BIC. The first ASW listed in
Table~\ref{table1}, with
$\omega_*L/(2\pi c)=0.4101$ for $a=0.4L$ and $\epsilon_1=8.2$, is even
in $x$. We calculate $F_e$ for this BIC as a function of $a$ with a
fixed $\epsilon_1=8.2$. The 
result is shown in Fig.~\ref{figone}(b). Since $F_e$ is real and
changes signs, it must have a zero. It turns out that $F_e=0$ for $a =
0.4404L$. The frequency of the corresponding ASW
is $\omega_* L/(2\pi c) = 0.3950$. Its wave field pattern
is shown in Fig.~\ref{fig_two}(a).

\begin{figure}[htb]
\centering 
\includegraphics[width=4.2cm]{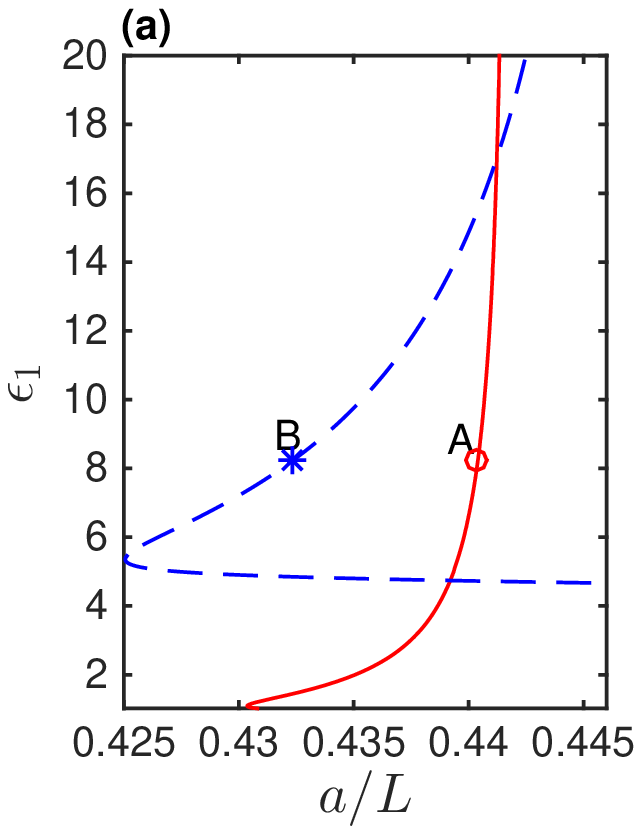}
\includegraphics[width=4.2cm]{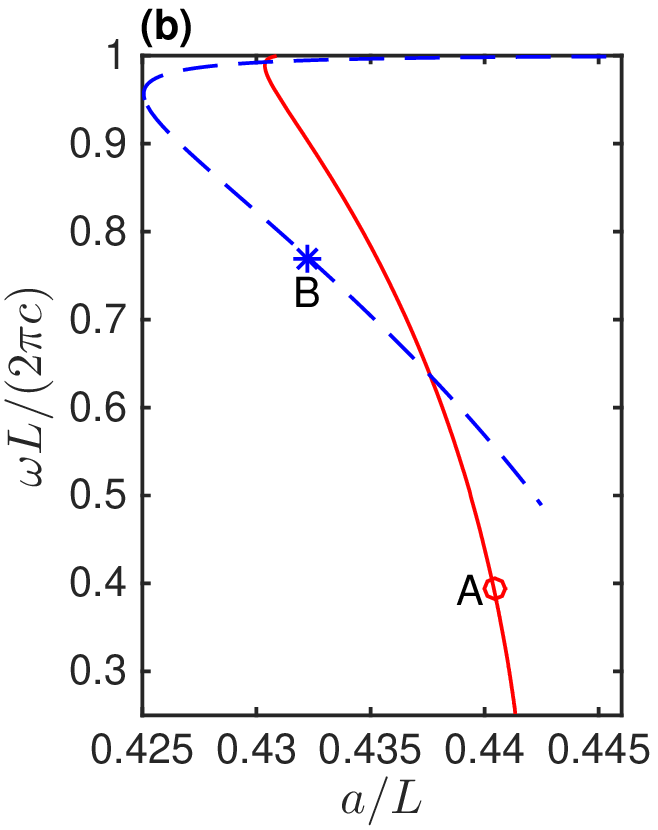}
\caption{(a): Parameters of the periodic array for 
  two special ASWs: the first ASW (red solid curve) and the fifth ASW
  (blue dashed curve). (b) The corresponding frequencies of the two special
  ASWs. Points ${\sf A}$ and ${\sf B}$ correspond to
  Figs.~\ref{fig_two}(a) and \ref{fig_two}(b), respectively.} 
\label{fig_three}
\end{figure}

For other values of $\epsilon_1 > 1$, $F_e$ of the first ASW
can still reach zero for a properly chosen $a$. Those values of $a$
and $\epsilon_1$ such that $F_e=0$ for the first ASW
give rise to a curve in the $a\epsilon_1$ plane, shown as the red
solid line in Fig.~\ref{fig_three}(a).
The corresponding frequency $\omega_*$ is shown with $a$ as the
solid red curve in Fig.~\ref{fig_three}(b). It appears that as
$\epsilon_1$ is increased, the related $a$ 
increases and approaches a constant as infinity. It
should be pointed that the first ASW exists
for all $\epsilon_1 > 1$ and $0 < a \le 0.5L$ \cite{hu15}. The curve
represents those parameter values such that the ASW becomes a
special BIC surrounded by strong resonances. 

For the other BICs listed in Table~\ref{table1}, we also attempt to find
parameters $a$ and $\epsilon_1$ such that $F_e=F_o=0$. It seems
that only the fifth ASW, with $\omega_*
L/(2\pi c) = 0.8749$ for $a=0.4L$ and $\epsilon_1=8.2$, can be tuned
to a special BIC. For
$\epsilon_1=8.2$, the fifth ASW, which is also even in $x$, gives $F_e=0$
for $a=0.4323L$. Its frequency is $\omega_* L/(2\pi c) = 0.7701$, and
its field pattern is shown in Fig.~\ref{fig_two}(b). For other values of
$\epsilon_1$, we also found the corresponding values of $a$ such that 
$F_e=0$  for the fifth ASW.  The results are given as a curve in the
$a\epsilon_1$ plane, i.e., the blue dashed line in
Fig.~\ref{fig_three}(a). The corresponding 
frequency $\omega_*$ is shown as the blue dashed line in
Fig.~\ref{fig_three}(b).  Notice that $\epsilon_1$ has a lower bound
around $4.67$, and it is achieved as $a \to 0.5L$. In addition, $a$
as a function of $\epsilon_1$, has a minimum around $\epsilon_1=5.35$,
and it seems to approach a constant as $\epsilon_1$ tends to infinity.

\begin{figure}[htb]
  \centering 
\includegraphics[scale=0.5]{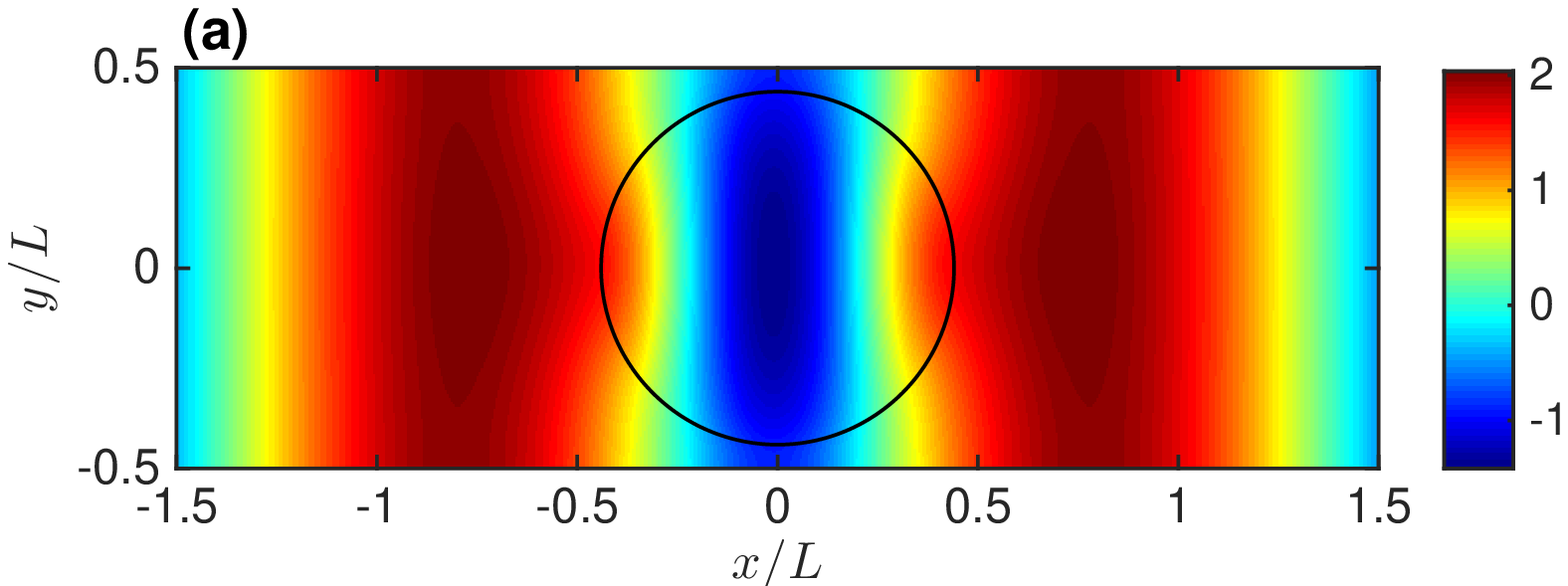}
\includegraphics[scale=0.5]{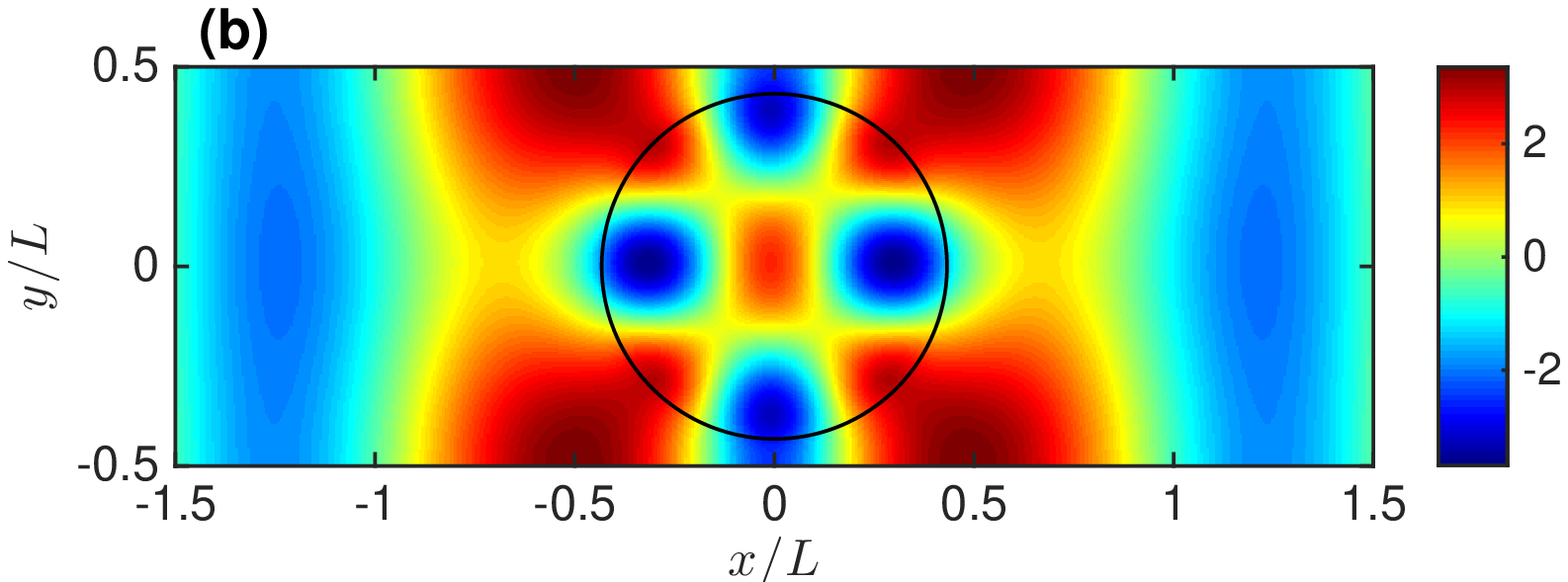}
  \caption{(a) and (b): Diffraction 
    solutions $\varphi_e$ corresponding to the special ASWs shown in
    Figs.~\ref{fig_two}(a) and \ref{fig_two}(b), respectively.} 
  \label{fig_four}
\end{figure}

In order to evaluate $F_e$ for an $x$-even BIC, we need to
calculate the $x$-even diffraction solution $u_e = C (u_l + u_r)$,
where $C$ is chosen so that $u_e$ is ${\cal PT}$-symmetric
and $C=e^{-i \tau}$ for a real constant $\tau$. As shown in
\cite{yuan17_4}, this leads to 
\begin{equation}
\varphi_e \sim 2 \cos(\alpha_* x \pm \tau), \quad x \to \pm \infty.  
\end{equation}
In Fig.~\ref{fig_four},
we show the diffraction solutions corresponding to the 
two special ASWs shown in Fig.~\ref{fig_two}.

\begin{figure}[htb]
  \centering
\includegraphics[width=8cm,height=5cm]{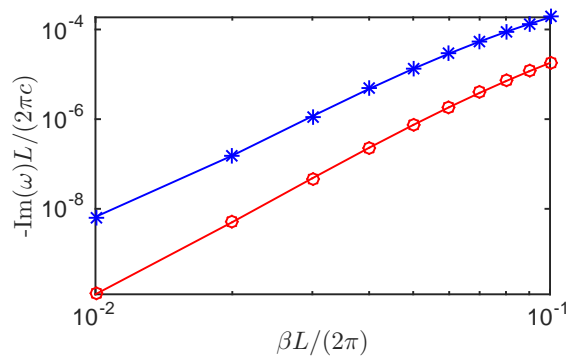}
\caption{Imaginary parts of the complex frequencies vs. $\beta$ for resonant modes near ASWs shown in
  Fig.~\ref{fig_two}(a) ($\circ$) and Fig.~\ref{fig_two}(b) ($\ast$), respectively.}
\label{fig_five}
\end{figure}

In Sec.~III, we argued that if $\mbox{Im}(\omega_2)=0$, then 
$\mbox{Im}(\omega_3)$ should also be zero, and 
$\mbox{Im}(\omega)$ should be proportional to
$(\beta-\beta_*)^4$ in general. For the two ASWs shown in 
Fig.~\ref{fig_two}, we check this result by computing the 
complex frequencies of some nearby resonant modes directly. 
In Fig.~\ref{fig_five},
we show $\mbox{Im}(\omega)$ as functions of $\beta$ in a logarithmic
scale for some resonant modes near these two ASWs. The numerical results
confirm the fourth order relation between $\mbox{Im}(\omega)$ and $\beta$.

\section{Conclusion}

BICs on periodic structures are surrounded by resonant modes with 
$Q$-factors approaching infinity. 
High-$Q$ resonances and the resulting strong local field enhancement
have important applications in lasing, nonlinear optics, etc. 
On 2D periodic structures, the $Q$-factors of the resonant modes near
a BIC usually satisfy $Q \sim 1/(\beta-\beta_*)^2$, where $\beta$ and
$\beta_*$ are the Bloch wavenumbers of the resonant mode and the BIC,
respectively. In this paper, we derived a general condition for
special BICs so that their nearby resonant modes have $Q \sim
1/(\beta-\beta_*)^4$.
These special BICs produce  high-$Q$ resonances for a very large range
of $\beta$, and they are useful because precise control of $\beta$
may be difficult in practice. The conditions for the special BICs are given in
integrals involving the BIC and related diffraction solutions, and they imply that the
first order perturbation $\phi_1$ does 
not radiate power to infinity. Numerical examples are given for two
families of ASWs on a periodic array of circular cylinders. 

In practical applications, the BICs always dissolve into resonant modes with
finite $Q$-factors, because the structures are always finite and
fabrication errors will break the required symmetries and
periodicity. We expect the special BICs have advantages over the
ordinary BICs in practical structures with fabrication errors and in
finite structures, but a rigorous analysis is still under
development. In addition, the results of this paper are restricted to 2D
structures. Clearly, it is worthwhile to derive similar conditions for
special BICs on bi-periodic three-dimensional (3D) structures and 
rotationally symmetric 3D structures.

\section*{Acknowledgments}
The authors acknowledge support from the Basic and Advanced Research
Project of CQ CSTC (Grant No. cstc2016jcyjA0491),  the Scientific and 
Technological Research Program of Chongqing Municipal Education
Commission (Grant No. KJ1706155), the Program for University Innovation Team of
Chongqing (Grant No. CXTDX201601026), and the Research Grants
Council of Hong Kong Special Administrative Region, China (Grant
No. CityU 11304117).

\section*{Appendix}

For operator ${\cal L}$ given in Eq.~(\ref{defL}), it is easy to
verify that 
\[
\overline{\phi}_* {\cal L} \phi_1
-  \phi_1 \overline{\cal L} \overline{\phi}_*
 = 
\nabla \cdot \left( \overline{\phi}_* \nabla \phi_1 
- \phi_1 \nabla \overline{\phi}_* \right)
+ 2 i \beta_* \partial_y (\phi_1 \overline{\phi}_*),
\]
where $\nabla$ is the 2D gradient operator. The integral on $\Omega$
of the right hand side can be reduced to an integral on $\partial
\Omega$ (the boundary of $\Omega$) by the divergence theorem. It
is zero, since $\phi_*$ and $\phi_1$ are periodic in $y$ and $\phi_*
\to 0$ exponentially as $|x| \to \infty$. Meanwhile, $\phi_*$
satisfies Eq.~(\ref{eq:phistar}), thus
$\int_\Omega \overline{\phi}_* {\cal L} \phi_1 d{\bm r} = 0$. 
From Eq.~(\ref{eq:phione}) for $\phi_1$, it is clear that 
\[
\int_\Omega \overline{\phi}_* \left[ - 2i \partial_y \phi_* +
  2(\beta_* - k_* k_1 \epsilon ) \phi_* \right] d{\bm r} = 0.
\]
This leads to Eq.~(\ref{eq:k1}). Meanwhile, $\int
\overline{\phi}_* \partial_y \phi_* d{\bm r}$ 
is pure imaginary, since 
\begin{eqnarray*}
\int_\Omega \overline{\phi}_* \partial_y \phi_* d{\bm r}
&=& \int_\Omega \partial_y |\phi_*|^2 d{\bm r} 
- \int_\Omega \phi_* \partial_y \overline{\phi}_* d{\bm r} \cr
&=& - \int_\Omega \phi_* \partial_y \overline{\phi}_* d{\bm r}.  
\end{eqnarray*}
Therefore, $k_1$ is real. 

Similarly, we have 
$\int_\Omega \overline{\phi}_* {\cal L} \phi_2 d{\bm r} =
0$. Multiplying both sides of Eq.~(\ref{eq:phitwo}) and integrating on
$\Omega$, we obtain
\[
k_2 = \frac{ \int_\Omega (1 - k_1^2 \epsilon) |\phi_*|^2 d{\bm r} +
  \int_\Omega R\, d{\bm r} }{ 2 k_* \int_\Omega \epsilon |\phi_*|^2 
  d{\bm r}},
\]
where $R = \overline{\phi}_* [ -2i \partial_y \phi_1 + 2(\beta_* - k_*
k_1 \epsilon) \phi_1]$. Therefore, 
\[
\mbox{Im}(k_2) = \frac{  \mbox{Im} \left[ \int_\Omega R\, d{\bm r} \right] }{ 2 k_*
  \int_\Omega \epsilon |\phi_*|^2    d{\bm r}}.
\]
From the complex conjugate of Eq.~(\ref{eq:phione}), we obtain
\[
\int_{\Omega_h} 
\phi_1 \overline{\cal L} \overline{\phi}_1  d{\bm r} 
= \int_{\Omega_h} R\, d{\bm r},
\]
where $\Omega_h$ is the rectangular domain defined in Sec.~IV. The
right hand side above requires an integration by parts that 
switches  the integral from $\phi_1 \partial_y \overline{\phi}_*$
to $- \overline{\phi}_* \partial_y \phi_1$. Meanwhile, it is easy to verify that 
\begin{eqnarray*}
&& \int_{\Omega_h} \phi_1 \overline{\cal L} \overline{\phi}_1  d{\bm r} 
= 
\int_{\partial \Omega_h} 
\phi_1 \frac{\partial \overline{\phi}_1}{\partial \nu} ds \cr
&& + \int_{\Omega_h} \left[  (k_*^2 \epsilon - \beta_*^2) |\phi_1|^2  -
   |\nabla \phi_1|^2  - 2i\beta_*   \phi_1 \frac{\partial
   \overline{\phi}_1}{\partial y} \right] d{\bm r}, 
\end{eqnarray*}
where $\partial \Omega_h$ is the boundary of $\Omega_h$ and $\nu$ is
its unit outward normal vector. The second term in the right hand side
above is real. Therefore, 
\[
\mbox{Im} \left[ \int_{\Omega_h} R\, d{\bm r} \right] 
=  \mbox{Im} \left[ 
\int_{\partial \Omega_h} 
\phi_1 \frac{\partial \overline{\phi}_1}{\partial \nu} ds \right].
\]
Since $\phi_1$ is periodic in $y$, the line integrals at $y=\pm L/2$ are
canceled. Therefore
\[
\int_{\partial \Omega_h} 
\phi_1 \frac{\partial \overline{\phi}_1}{\partial \nu} ds
= 
\int_{-L/2}^{L/2} \left[ 
 \phi_1 \frac{\partial \overline{\phi}_1}{\partial x}
\right]^{x=h}_{x=-h} dy,
\]
where $P(x,y)|_{x=-h}^{x=h}$ denotes $P(h,y)-P(-h,y)$. 

For $|x| > D$, the equation for $\phi_1$ is quite simple. It is not
difficult to see that 
\[
\phi_1 = d_0^\pm e^{ \pm i \alpha_* x} 
+ \sum_{j\ne 0} d_j^\pm(x) e^{ i 2\pi jy/L} e^{\pm \gamma_{*j} x}
\]
for $x> D$ and $x< -D$ respectively, where $\gamma_{*j} = -i
\alpha_{*j}$ is positive, 
$d_0^\pm$ are unknown coefficients, and $d_j^\pm(x)$ ($j\ne 0$) are unknown
linear polynomials of $x$. The above gives
\[
\lim_{h \to +\infty} \int_{-L/2}^{L/2} \left[ 
 \phi_1 \frac{\partial \overline{\phi}_1}{\partial x}
\right]^{x=h}_{x=-h} dy 
= -i L \alpha_* (|d_0^+|^2 + |d_0^-|^2), 
\]
and $\mbox{Im} \left[ \int_\Omega R\, d{\bm r} \right] = - L \alpha_*
(|d_0^+|^2 +   |d_0^-|^2)$, and finally Eq.~(\ref{eq:k2_imag}). 

To show Eq.~(\ref{leftinc}), we notice that 
\[
\overline{\varphi}_l {\cal L} \phi_1 
-\phi_1 \overline{\cal L} \overline{\varphi}_l = 
\nabla \cdot [ \overline{\varphi}_l \nabla \phi_1 
- \phi_1 \nabla \overline{\varphi}_l ] 
+ 2i \beta_* \partial_y (
\phi_1 \overline{\varphi}_l).
\]
Since $\varphi_l$ satisfies the Helmholtz equation and both $\phi_1$
and $\varphi_l$ are periodic in $y$, we have 
\[
\int_{\Omega_h} \overline{\varphi}_l {\cal L} \phi_1 d{\bm r}
= 
\int_{\partial \Omega_h} \left[ \overline{\varphi}_l \frac{ \partial
    \phi_1 }{\partial \nu} 
- \phi_1 \frac{ \partial \overline{\varphi}_l }{\partial \nu} \right] ds. 
\]
In the right hand side above, the integrals on the two edges at
$y=\pm L/2$ are canceled. Therefore 
\[
\int_{\Omega_h} \overline{\varphi}_l {\cal L} \phi_1 d{\bm r}
= 
\int_{-L/2}^{L/2}  \left[ \overline{\varphi}_l \frac{ \partial
    \phi_1}{\partial x}
- \phi_1 \frac{ \partial \overline{\varphi}_l }{\partial x} \right]^{x=h}_{x=-h} dy.
\]
Based on the asymptotic formula (\ref{leftasym}), it is easy to show
that as $h \to +\infty$, the right hand side above tend to $2i L
\alpha_* (d_0^+ \overline{T}_l + d_0^- \overline{R}_l)$. This leads to
Eq.~(\ref{leftinc}). The proof for Eq.~(\ref{rightinc}) is similar.

\end{document}